\documentstyle[preprint,aps,prl]{revtex}
\begin{document}
\widetext
\title{
Phase diagram of the bose Hubbard model
}
\author{
J. K. Freericks\cite{Davis} and H. Monien\cite{ETH}.
}
\address{
Institute for Theoretical Physics,
University of California,
Santa Barbara, CA 93106
}
\date{July 7, 1993}
\maketitle
\widetext
\begin{abstract}
The first reliable analytic calculation
of the phase diagram of the bose gas on a $d$-dimensional
lattice with on-site repulsion is presented.
In one dimension, the analytic calculation is in excellent agreement
with the numerical Monte Carlo results.
In higher dimensions, the deviations
from the Monte Carlo calculations are larger,
but the
correct shape of the Mott insulator lobes
is still obtained. Explicit expressions
for the energy of the Mott and the ``defect'' phase
are given in a strong-coupling expansion.
\end{abstract}
\narrowtext
Strongly interacting quantum systems continue to challenge
theoretical physics.
Fermionic examples of strong correlations include
high temperature superconductors,
heavy fermion materials and metal-insulator
transitions.
The study of these systems has been confined to either
numerical simulations which are plagued by finite-size effects
(with the notable exception of the infinite-dimensional expansion)
or uncontrollable approximations.
Similar problems arise in
strongly interacting bosonic systems which have attracted
a lot of recent interest \cite{Fishers,Scalettar,Rokhsar,Girvin}.
Physical realizations
include short correlation length superconductors, granular
superconductors, Josephson arrays, and the dynamics of flux lattices
in type II superconductors.
The relevant physics of these problems is contained in the bose Hubbard
Hamiltonian which describes the competition between kinetic
energy and potential energy effects.
Various aspects of this model were investigated analytically by
mean-field theory \cite{Fishers,KampfZimanyi}, by renormalization group
techniques \cite{Fishers,Rokhsar} and by projection methods \cite{Krauth0}.
The bose Hubbard model was also studied with Quantum Monte
Carlo methods (QMC) by Batrouni et al.  \cite{Scalettar}
in one dimension (1+1)
and by Krauth and Trivedi \cite{Krauth3} in two dimensions (2+1).
In this contribution, we show that the phase diagram
obtained from
a strong-coupling expansion has the correct dependence on the dimensionality
of the spatial lattice, and agrees with the
QMC calculations (to within a few percent).

We study the minimal model which contains the key physics of the strongly
interacting bose system -- the competition between kinetic and potential
energy effects:
\begin{equation}
  H = - \sum_{ij} t_{ij} b^\dagger_i b^{\phantom{\dagger}}_j
   - \mu \sum_i {\hat n_i} + \frac{1}{2} U \sum_i {\hat n}_i ({\hat n}_i-1)
  \quad , \quad
  {\hat n}_i =  b^\dagger_i b^{\phantom{\dagger}}_i
  \label{H}
\end{equation}
where $b_i$ is the boson
annihilation operator at site $i$, $t_{ij}$ is the hopping matrix element
between the site $i$ and $j$,
$U$ is the strength of the on-site repulsion, and $\mu$ is the chemical
potential.
The approximate form of the zero temperature ($T=0$)
phase diagram can be understood
by starting from the strong-coupling or ``atomic'' limit
\cite{Fishers,GiamarchiSchulz,Ma}.
In this limit, the kinetic energy vanishes ($t_{ij} = 0$)
and every site is occupied by a fixed
number of bosons, $n_0$.
The ground-state boson occupancy ($n_0$)
is then chosen in
such a way as to minimize the on-site energy.
If the chemical potential, $\mu=(n_0+\delta)U$, is parametrized
in terms of the deviation, $\delta$, from
integer filling $n_0$, then the on-site energy is
$E(n_0) = -\delta U n_0 - \frac{1}{2} U n_0 (n_0 - 1)$, and
the energy to add a boson
onto a particular site satisfies $E(n_0+1) - E(n_0) = -\delta U n_0$.
Thus for a nonzero $\delta$, a finite amount of energy (gap)
is required to
move a particle through the lattice.
The bosons are localized, producing
a Mott insulator. This energy gap decreases with increasing
strength of the hopping matrix elements until it
vanishes and the bosons condense into the superfluid phase.
For $\delta = 0$ the energy of the two different boson densities is degenerate
[$E(n_0) = E(n_0+1)$] and no energy is needed
to add or extract a particle;
i.e., the compressibility is finite.
As the strength of the hopping matrix elements increases, the range of
the chemical potential $\mu$ about
which the system is incompressible decreases.
The Mott-insulator phase will completely disappear at a critical value of the
hopping matrix elements.

We are interested in the determination of the phase boundary
between the (incompressible) Mott insulator and the (compressible)
superfluid phase.
A strong-coupling expansion for this boundary is determined by calculating both
the energy of the Mott insulating phase and of a defect phase (which contains
an extra hole or particle) in a perturbative expansion of the kinetic energy.
At the point where the energy of the Mott phase is degenerate with the defect
phase, the system becomes compressible, and, since there is no disorder,
also superfluid \cite{Fishers}.
There are two distinct cases for the defect phase:
$\delta < 0$ corresponds to adding a boson to the Mott-insulator phase (with
$n_0$ bosons per site); and $\delta>0$ corresponds to adding a hole to the
Mott-insulator phase (with $n_0+1$) bosons per site.
The phase boundary will depend on the number of bosons per site, $n_0$, of the
initial Mott insulator phase.

To zeroth order in $t/U$ the Mott insulating state is given by
\begin{equation}
  |\Psi_{\text{Mott}}(n_0)\rangle^{(0)} =
  \prod_{i=1}^N
  \frac{1}{\sqrt{n_0!}}\left(b^\dagger_i\right)^{n_0}|0\rangle
\end{equation}
where $n_0$ is the number of bosons on each site, $N$ is the number of
sites in the lattice and $|0\rangle$ is the vacuum state.
The defect phase is characterized by one additional particle (hole) which moves
coherently throughout the lattice. To zeroth order in $t/U$
the wave function for the ``defect phase'' is determined by degenerate
perturbation theory:
\begin{eqnarray}
  |\Psi_{\text{Def}}(n_0)\rangle^{(0)}_{\text{particle}} &=&
    \frac{1}{\sqrt{n_0+1}} \sum_i f_i b^\dagger_i
    |\Psi_{\text{Mott}}(n_0)\rangle^{(0)}
    \cr
  |\Psi_{\text{Def}}(n_0)\rangle^{(0)}_{\text{hole}} &=&
    \frac{1}{\sqrt{n_0}} \sum_i f_i b^{\phantom{\dagger}}_i
    |\Psi_{\text{Mott}}(n_0)\rangle^{(0)}
\end{eqnarray}
where the $f_i$ is the eigenvector of the hopping matrix $t_{ij}$ with the
lowest eigenvalue \cite{NondegenerateAssumption}.

For simplicity we will
only consider hopping between the nearest neighbors of a hypercubic
lattice in $d$-dimensions.
The number of nearest neighbors is denoted by $z=2d$ and the hopping matrix
element by $t$; the minimum eigenvalue of the hopping matrix is $-zt$.
The many body version of standard Rayleigh-Schr\"odinger
perturbation theory is employed throughout.
To third order in $t/U$, the energy of the Mott state with
$n_0$ bosons per site becomes
\begin{equation}
  E_{\text{Mott}}(n_0) = N
  \left[
  -\delta U n_0 - \frac{1}{2}U n_0 (n_0+1) - \frac{zt^2}{U}n_0(n_0+1)
  \right]
  \label{eq: EMott}
\end{equation}
which is proportional to the number of sites. Note that the odd-order
terms vanish.
The energy difference between the Mott insulating phase
and the defect phase with
an additional particle ($\delta < 0$) satisfies
\begin{eqnarray}
  E_{\text{Def}}^{(particle)}(n_0) - E_{\text{Mott}}(n_0)
  &=&
  -\delta^{(\text{particle})} U-zt(n_0+1)
  + \frac{zt^2}{U}\frac{n_0(5n_0+4)}{2}
  -\frac{z^2t^2}{U}n_0(n_0+1)\cr
  &+&
  \frac{t^3}{U^2} n_0(n_0+1)
  \left[(-2z^3+\frac{25}{4}z^2-4z)n_0 + (-z^3+\frac{7}{2}z^2-2z)\right]
  \label{eq: Edef upper}
\end{eqnarray}
to third order in $t/U$;
while the energy difference
between the Mott insulating phase and
the defect phase with an additional hole ($\delta > 0$) satisfies
\begin{eqnarray}
  E_{\text{Def}}^{(hole)}(n_0) - E_{\text{Mott}}(n_0)
  &=&
  \delta^{(\text{hole})}U-zt n_0
  + \frac{zt^2}{U}\frac{(n_0+1)(5n_0+1)}{2}
  -\frac{z^2t^2}{U}n_0(n_0+1)\cr
  &+&
  \frac{t^3}{U^2} n_0(n_0+1)
  \left[(-2z^3+\frac{25}{4}z^2-4z)n_0 + (-z^3+\frac{11}{4}z^2-2z)\right]
  \label{eq: Edef lower}
\end{eqnarray}
These results have been verified by small-cluster calculations
on two and four-site clusters.
Note that the energy difference in Eqs. (\ref{eq: Edef upper}) and (\ref{eq:
Edef lower})
is {\em independent} of the lattice
size $N$.

The phase boundary between the incompressible Mott phase and the compressible
superfluid phase occurs when the
energy difference between the two different phases vanishes
\cite{ContinousAssumption}.
The two branches of the Mott phase boundary meet when
\begin{equation}
  \delta^{(particle)}(n_0) + 1 = \delta^{(hole)}(n_0).
  \label{eq: critical condition}
\end{equation}
The additional one on the left hand side
arises because $\delta$ is measured from the point
$\mu/U = n_0$.
Equation (\ref{eq: critical condition})
may be used to estimate the critical value of
the hopping matrix element, $t_{critical}(n_0)$, beyond which no Mott-insulator
phase exists.

In one dimension,
the upper boundary of the Mott insulator lobe (with a particle density
of $n_0$) is given by
\begin{equation}
  \delta^{(particle)}(n_0, t/U)
  = - 2 ( n_0 + 1) (t/U) + n_0^2 (t/U)^2 + n_0 (n_0+1)(n_0+2) (t/U)^3
  \label{eq: upper boundary}
\end{equation}
to third order in $t/U$,
and the lower boundary is given by
\begin{equation}
  \delta^{(hole)}(n_0, t/U)
  =  2 n_0 (t/U)  - (n_0+1)^2 (t/U)^2 + n_0 (n_0+1)(n_0-1) (t/U)^3.
  \label{eq: lower boundary}
\end{equation}
The slope of the phase boundaries about the point $\mu=n_0 U$ are equal in
magnitude to first order
$
[
\lim_{t\rightarrow 0}\frac{d}{dt}\delta^{particle}(n_0,t/U) =
- \lim_{t\rightarrow 0}\frac{d}{dt}\delta^{hole}(n_0+1,t/U)
]
$,
but change in magnitude as a function of the density ($n_0$),
implying that the
Mott-phase lobes always have an asymmetrical shape.

The strong-coupling expansion for the $t$, $\mu$
phase diagram in one dimension is compared to the
QMC results of Batrouni et al. \cite{Scalettar}
in Figure 1.
The solid lines indicate the phase boundary between the Mott-insulator phase
and the superfluid phase at zero temperature
as calculated from Eq.~(\ref{eq: upper boundary})
and Eq.~(\ref{eq: lower boundary}).
The squares are the results of the QMC calculation at
a small but finite temperature $T = U/2$ \cite{Scalettar}.
Note that the overall agreement of the two
calculations is excellent.
For example, the critical value of the chemical potential for the first lobe
($n_0=1$) satisfies $\delta_{critical} \approx -0.755$ so that
the critical value of the hopping matrix element is
$(t/U)_{critical} = 0.215$, while the QMC calculations
found $(t/U)_{critical} = 0.215 \pm 0.01$ \cite{Scalettar}.
A closer examination shows that the first
lobe ($n_0 = 1$) has a systematic deviation at larger values of $t$.
This is most likely a finite-temperature effect, since the Mott-insulator
phase becomes more stable at higher temperatures \cite{KampfZimanyi}.

It is known from the scaling theory of Fisher et al. \cite{Fishers} that the
phase transition at the tip of the Mott lobe is in the universality class
of the $(d+1)$ dimensional $XY$ model.
Although a finite-order perturbation theory cannot describe the physics of
the tricritical point correctly, it turns out that the density fluctuations
dominate the physics of the phase transition even close to the tricritical
point.
Note how the Mott lobes have a cusp-like structure in one dimension, mimicking
the Kosterlitz-Thouless behavior of the critical point.

Figure 2 presents the strong-coupling expansion for the
$t$, $\mu$ phase diagram in two dimensions.
For comparison, the tricritical point of the first
Mott-insulator lobe as obtained by
the QMC simulations of
Krauth and Trivedi \cite{Krauth3} is marked with a solid square.
Their numerical calculation gives
a critical value of $(t/U)_{critical} = 0.122\pm0.01$,
whereas our calculation yields $(t/U)_{critical} \approx 0.136$
which is in reasonable agreement.
As already mentioned above we cannot hope to describe the physics close to
tricritical point with our approach, but note that the qualitative shape
of the Mott lobes has changed from one dimension to two dimensions,
mimicking the ``smooth'' critical behavior of the $XY$ model
in three or larger dimensions.

Finally the strong-coupling expansion is compared to the exact calculation
in infinite dimensions \cite{Fishers}.
In infinite dimensions, the hopping matrix element must scale inversely with
the dimension \cite{MuellerHartmann}, $t=t^*/d$, $t^* = \text{finite}$,
producing the mean field theory of Ref. \cite{Fishers}.
In Figure 3 the strong-coupling expansion
(solid line) is compared to the exact solution (dashed line).
Even in infinite dimensions, the agreement of the strong-coupling expansion
with the exact results is quite good.

We have repeatedly compared a strong-coupling expansion
to the numerical QMC simulations
for the incompressible-compressible phase boundary of the bose Hubbard model.
A mean-field treatment of the bose Hubbard model
(e.g. \cite{Fishers,KampfZimanyi}) cannot
capture the physics of the one dimensional system which is completely
dominated by fluctuations. The dimensionality only enters as a trivial
prefactor in integrals over the phase space. For this reason, mean-field
theories will always give a concave shape to the Mott-insulator lobes
independent of the dimension.
A strong-coupling expansion, on the other hand, easily distinguishes the shape
difference from one dimension to higher dimensions and shows that a proper
treatment of density fluctuations is critical in determining the Mott-insulator
to superfluid transition.
In conclusion we have described an analytical method to accurately
calculate the phase diagram
of the bose Hubbard model in any dimension.
Extensions of these techniques to include disorder will be
presented separately.

\acknowledgments
We would like to thank R. Scalettar, G. Batrouni and K. Singh,
for providing us with the Quantum Monte Carlo
data for the one dimensional bose Hubbard model and
for many useful discussions.
This research was supported in part by the NSF under Grant No.
PHY89-04035 and DMR90-02492.

\begin{figure}[t]
  \caption{
  The $t$, $\mu$ phase diagram of the bose Hubbard model in one
  dimension ($d = 1$). The solid lines give the phase boundaries of the
  Mott insulator to the superfluid state as determined from
  a third-order strong-coupling calculation.
  The squares are the result of the QMC
  calculation of Batrouni et al. [2].
  }
  \label{fig:1}
\end{figure}
\begin{figure}[t]
  \caption{
  The $t$, $\mu$ phase diagram of the bose Hubbard model in two
  dimensions ($d = 2$). The solid lines give the phase boundaries of the
  Mott insulator to the superfluid state as determined from
  a third-order strong-coupling calculation.
  The point indicates the tricritical point as
  determined by the QMC calculation of Krauth and
  Trivedi [7].
  }
  \label{fig:2}
\end{figure}
\begin{figure}[t]
  \caption{
  The $t$, $\mu$ phase diagram of the bose Hubbard model in infinite
  dimensions ($d \rightarrow \infty$).
  The solid lines give the phase boundaries of the
  Mott insulator to the superfluid state as determined from
  a third-order strong-coupling calculation.
  The dashed lines are the result of the mean field
  calculation of Fisher et al. [1].
  }
  \label{fig:3}
\end{figure}


\begin{thebibliography}{10}
\bibitem{NondegenerateAssumption}
The hopping matrix is assumed to have a nondegenerate lowest eigenvalue.
\bibitem{ContinousAssumption}
The compressibility is assumed to vanish continuously at the phase boundary.
\end{thebibliography}
\end{document}